\pdfoutput=1

\documentclass[twocolumn,twoside,nofootinbib,showpacs,prd,aps,superscriptaddress,tightenlines,preprintnumbers,10pt]{revtex4-1}

\usepackage{amsmath}
\usepackage{slashed}
\usepackage{graphicx}
\usepackage[hyperfootnotes=false,bookmarks=false]{hyperref}
\usepackage{amssymb}
\usepackage{mathrsfs}
\usepackage{bm}
\usepackage{xspace}
\usepackage{booktabs}
\usepackage{xcolor}

\newcommand{\eq}[1]{Eq.~\eqref{eq:#1}}
\newcommand{\eqs}[2]{Eqs.~\eqref{eq:#1} and \eqref{eq:#2}}
\renewcommand{\sec}[1]{Sec.~\ref{sec:#1}}

\newcommand{\citere}[1]{Ref.~\cite{#1}}
\newcommand{\tab}[1]{Table~\ref{tab:#1}}

\newcommand{\ord}[1]{{\mathcal O}(#1)}
\newcommand{\ORd}[1]{{\mathcal O}\Bigl(#1\Bigr)}

\newcommand{\BI}{\begin{itemize}}
\newcommand{\EI}{\end{itemize}}
\newcommand{\BE}{\begin{equation}}
\newcommand{\EE}{\end{equation}}
\newcommand{\BEA}{\begin{eqnarray}}
\newcommand{\EEA}{\end{eqnarray}}

\newcommand{\nn}{\nonumber}

\newcommand{\df}{\mathrm{d}}

\newcommand{\lra}{\leftrightarrow}

\newcommand{\al}{\alpha}

\newcommand{\ga}{\gamma}
\newcommand{\Ga}{\Gamma}
\newcommand{\de}{\delta}

\newcommand{\si}{\sigma}
\newcommand{\w}{\omega}

\newcommand{\cI}{{\mathcal I}}

\newcommand{\cL}{{\mathcal L}}

\newcommand{\cS}{{\mathscr S}}

\newcommand{\bn}{{\bar{n}}}

\newcommand{\lqcd}{\Lambda_\mathrm{QCD}}

\newcommand{\CS}{\cS}

\newcommand{\Ecm}{E_\mathrm{cm}}

\newcommand{\SCETp}{\ensuremath{{\rm SCET}_+}\xspace}

\allowdisplaybreaks[2]

\begin{document}

\preprint{\vbox{\hbox{NIKHEF 2016-016}}}

\title{Joint transverse momentum and threshold resummation beyond NLL}

\author{Gillian~Lustermans}
\affiliation{Nikhef, Theory Group, Science Park 105, 1098 XG, Amsterdam, The Netherlands}
\affiliation{ITFA, University of Amsterdam, Science Park 904, 1018 XE, Amsterdam, The Netherlands}

\author{Wouter J.~Waalewijn}
\affiliation{Nikhef, Theory Group, Science Park 105, 1098 XG, Amsterdam, The Netherlands}
\affiliation{ITFA, University of Amsterdam, Science Park 904, 1018 XE, Amsterdam, The Netherlands}

\author{Lisa Zeune}
\affiliation{Nikhef, Theory Group, Science Park 105, 1098 XG, Amsterdam, The Netherlands}

\begin{abstract}
To describe the transverse momentum spectrum of heavy color-singlet production, the joint resummation of threshold and transverse momentum logarithms is investigated. We obtain factorization theorems for various kinematic regimes valid to all orders in the strong coupling, using Soft-Collinear Effective Theory. We discuss how these enable resummation and how to combine regimes. The new ingredients in the factorization theorems are calculated at next-to-leading order, and a range of consistency checks is performed. Our framework goes beyond the current next-to-leading logarithmic accuracy (NLL).
\end{abstract}

\maketitle

\section{Introduction}\label{sec:intro}

In heavy particle production the additional radiation tends to be soft, due to the steeply falling parton distribution functions (PDFs). This implies that threshold logarithms of $1-z \equiv 1-Q^2/\hat s$ in the partonic cross section are large, where $Q$ is the heavy particle invariant mass and $\hat s$ the partonic center-of-mass energy. The corresponding threshold resummation can significantly modify the cross section. Well-known examples are top-quark pair production or the production of supersymmetric particles. When the $p_T$ of the heavy particle(s) is parametrically smaller than $Q$, the transverse momentum resummation of the logarithms of $p_T/Q$ is important as well.

In this letter we study the joint resummation of threshold and transverse momentum logarithms. A  formalism that achieves this at next-to-leading logarithmic (NLL) order has been developed some time ago~\cite{Laenen:2000ij} (see also \citere{Li:1998is}). Here resummation is simultaneously performed in Mellin moment (of $z$) and impact parameter (Fourier conjugate to $p_T$), accounting for the recoil of soft gluons using non-Abelian exponentiation and including the recoil in the kinematics of the hard scattering. This framework has been applied to prompt-photon~\cite{Laenen:2000de}, electroweak~\cite{Kulesza:2002rh}, Higgs boson~\cite{Kulesza:2003wn}, heavy-quark~\cite{Banfi:2004xa}, slepton pair~\cite{Bozzi:2007tea} and gaugino pair~\cite{Debove:2011xj} production.

We introduce a framework for joint resummation using Soft-Collinear Effective Theory (SCET)~\cite{Bauer:2000ew,Bauer:2000yr,Bauer:2001ct,Bauer:2001yt,Bauer:2002nz,Beneke:2002ph}, which enables us to go beyond NLL.  We need to assume a relative power counting between the threshold parameter $1\!-\!z$ and transverse momentum $p_T \gg \lqcd$ to derive factorization theorems, and identify the following three regimes
\begin{enumerate}
\item $1 \sim 1-z \gg p_T/Q$: transverse mom.~factorization
\item $1 \gg 1-z \gg p_T/Q$: \!intermediate regime
\item $1 \gg 1-z \sim p_T/Q$: threshold factorization
\end{enumerate}
The factorization theorems for regimes 1 and 3 are simply a more differential version of the standard transverse momentum and threshold resummation. The intermediate regime 2 requires us to extend SCET with additional collinear-soft (csoft) degrees of freedom. Such theories, typically referred to as \SCETp, have recently been used to describe a range of joint resummations~\cite{Bauer:2011uc, Procura:2014cba, Larkoski:2015zka, Larkoski:2015kga,Becher:2015hka,Chien:2015cka,Pietrulewicz:2016nwo}. We will elaborate on how the factorization in SCET leads to resummation using the renormalization group (RG) evolution. As a byproduct, this implies an all-order relation between the anomalous dimension of the thrust soft function and threshold soft function. We discuss how to combine the different factorization theorems describing the three regimes, finding that regime 2 can be obtained from regime 1 by a proper modification of renormalization scales, but that regime 3 contains additional corrections beyond NLL. By using SCET, gauge invariance is manifest, and  the ingredients in factorization theorems have matrix element definitions.
We will focus on the production of a color neutral state $pp \to V+X$ with $V =Z, W, h, \dots$, working in momentum space. All ingredients will be collected for joint resummation at next-to-next-to-leading logarithmic order (NNLL).

This letter is organized as follows. In~\sec{factor} we present the factorization theorem for joint resummation in each regime, and derive consistency relations between them. All ingredients entering the factorization formula are collected at next-to-leading order (NLO) in \sec{1loop}, and the consistency between regimes is verified. The renormalization group equations are given in \sec{anom}, providing an internal consistency check on each individual regime. In \sec{resummation} we discuss how to perform the resummation and combine the cross section in the three regimes. 
We conclude in~\sec{conclusions}.

\section{Factorization}\label{sec:factor}

In this section we present the factorization theorems that enable the joint resummation of threshold and transverse momentum logarithms, address a subtlety that arises at partonic threshold, and derive consistency relations between the regimes.

\subsection{Factorization theorems}\label{sec:factor1}

\begin{figure*}
 \includegraphics[width=\textwidth]{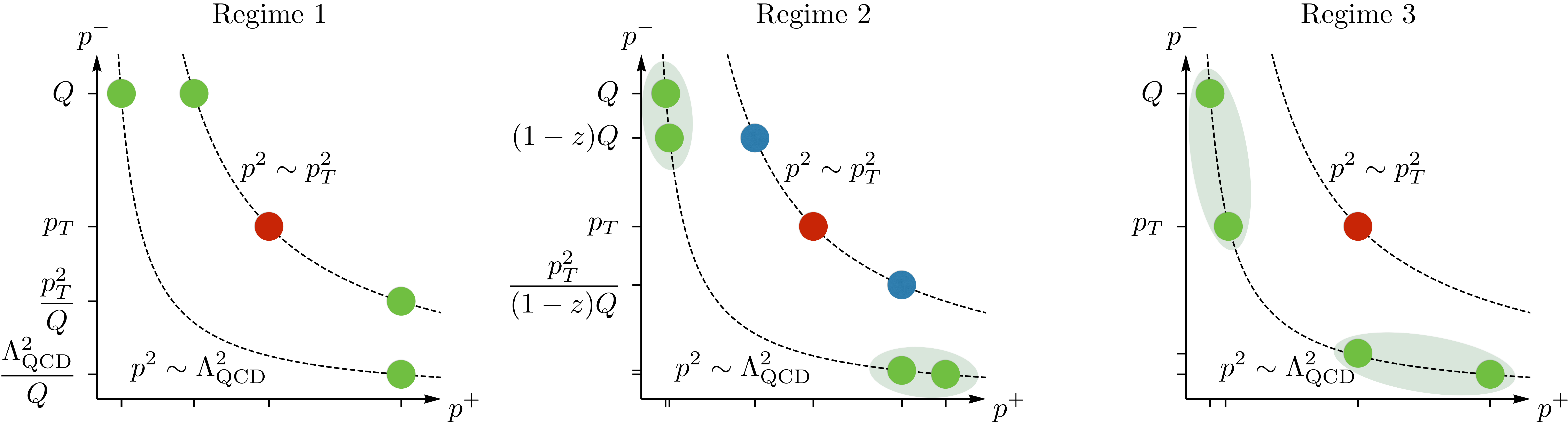}
 \caption{The power counting of collinear (green), collinear-soft (blue) and soft (red) modes for the three regimes. The dashed lines show invariant mass hyperbolas, and the nonperturbative modes that we do not factorize are grouped together.}
 \label{fig:modes}
\end{figure*}

\begin{table*}
\centering
   \begin{tabular}{l|ccc} 
     \hline \hline
     Regime: & 1: $Q \sim (1-z)Q \gg p_T \gg \lqcd$ & 2: $Q \gg (1-z)Q \gg p_T \gg \lqcd$ & 3: $Q \sim (1-z)Q \sim p_T \gg \lqcd$ \\
     \hline
     $n$-collinear & $(\lqcd^2/Q,Q,\lqcd)$ &  $(\lqcd^2/Q,Q,\lqcd)$ & $(\lqcd^2/Q,Q,\lqcd)$ \\
      & $(p_T^2/Q,Q,p_T)$ &  $\bigl(\lqcd^2/[(1-z)Q],(1-z)Q,\lqcd\bigr)$ & $\bigl(\lqcd^2/[(1-z)Q],(1-z)Q,\lqcd\bigr)$ \\
     $\bn$-collinear & $(Q,\lqcd^2/Q,\lqcd)$ & $(Q,\lqcd^2/Q,\lqcd)$ & $(Q,\lqcd^2/Q,\lqcd)$\\
      &$(Q,p_T^2/Q,p_T)$ & $\quad \bigl((1-z)Q,\lqcd^2/[(1-z)Q],\lqcd\bigr)\quad$ & $\bigl((1-z)Q,\lqcd^2/[(1-z)Q],\lqcd\bigr)$\\
     $n$-csoft & & $\bigl(p_T^2/[(1-z)Q], (1-z) Q, p_T\bigr)$ \\
     $\bn$-csoft & & $\bigl((1-z) Q, p_T^2/[(1-z)Q], p_T\bigr)$ \\     
     soft & $(p_T,p_T,p_T)$ & $(p_T,p_T,p_T)$ &$(p_T,p_T,p_T)$\\
     \hline \hline
   \end{tabular} 
   \caption{The modes (rows) and power counting of momenta in light-cone coordinates \eq{lc}, for the three regimes (columns).}
    \label{tab:modes1}
\end{table*}

In the introduction we identified three kinematic regimes, depending on the relative power counting of the transverse momentum and threshold parameter. 
The corresponding modes are shown in Fig. \ref{fig:modes} and are summarized in \tab{modes1}, using light-cone coordinates
\begin{align} \label{eq:lc}
  p^\mu = (p^+, p^-, p_\perp^\mu) = (p^0 - p^3, p^0 + p^3, p_\perp^\mu)
\,.\end{align}
We illustrate the origin of these degrees of freedom for regime 2. The incoming proton in the $n^\mu=(1,0,0,1)$ direction is described by a mode whose momentum components have the parametric size $(p^+, p^-, p_\perp^\mu) \sim (\lqcd^2/Q,Q,\lqcd)$. This scaling is fixed by its energy $\sim Q$ and virtuality $\sim \lqcd$.
Since we are in the threshold limit, the energy of the real radiation is $\lesssim (1-z)Q$. Collinear splittings within the proton therefore require an additional mode $(\lqcd^2/[(1-z)Q],(1-z)Q,\lqcd)$. It is natural to combine these nonperturbative modes into the (threshold) PDF, as they are both required for the PDF evolution.
The (isotropic) soft radiation that contributes to the $p_T$ measurement has the parametric scaling $(p_T,p_T,p_T)$. In addition there are collinear-soft (csoft) modes with scaling $(p_T^2/[(1-z)Q], (1-z) Q, p_T)$, which are uniquely fixed by their sensitivity to the transverse momentum measurement and threshold restriction~\cite{Procura:2014cba}. Similarly there are collinear and csoft modes in the $\bar{n}^\mu = (1,0,0,-1)$ direction.

This leads to the following factorization theorems that hold to all orders in the strong coupling,
\begin{widetext}
\begin{align} 
\text{Regime 1:} \quad  \frac{\df \si_1}{\df Q^2\, \df p_T} &= \si_0 \sum_{i,j} H_{ij}(Q^2,\mu) 
\int\! \df^2 \vec p_{a\perp}\,\df^2 \vec p_{b\perp}\,\df^2 \vec p_{s\perp}\, \de\bigl(p_T - |\vec p_{a\perp} + \vec p_{b\perp} + \vec p_{s\perp} |\bigr)
\nn  \\ & \quad \times
\int\! \df x_a\, \df x_b\, \de(\tau - x_a x_b)\,
B_i(x_a, \vec p_{a\perp},\mu,\nu)\,B_j(x_b,\vec p_{b\perp},\mu,\nu)\, S_{ij}(\vec p_{s\perp},\mu,\nu)
\,,\\  
\text{Regime 2:} \quad \frac{\df \si_2}{\df Q^2\, \df p_T} &= \si_0 \sum_{i,j} H_{ij}(Q^2,\mu) 
\int\! \df^2 \vec p^{\rm \,\,cs}_{a\perp}\,\df^2 \vec p^{\rm \,\,cs}_{b\perp}\,\df^2 \vec p_{s\perp}\,   
\de\bigl(p_T - |\vec p^{\rm \,\,cs}_{a\perp} + \vec p^{\rm \,\,cs}_{b\perp} + \vec p_{s\perp} |\bigr)
\nn \\ & \quad \times
\int\! \frac{\df \xi_a}{\xi_a}\, \frac{\df \xi_b}{\xi_b}\, \df p^{\rm cs\,-}_a\, \df p^{\rm cs\,+}_b\,
\de\Bigl[1 - \frac{\tau}{\xi_a \xi_b} - \Bigl(\frac{p_a^{\rm cs\,-}}{Q} + \frac{p_b^{\rm cs\,+}}{Q}\Bigr) \Bigr]\,  
f_i(\xi_a,\mu)\, f_j(\xi_b, \mu)\, 
\nn \\ & \quad \times
\cS_i\bigl(p_a^{\rm cs\,-},\vec p^{\rm \,\,cs}_{a\perp},\mu,\nu\bigr)\,
\cS_j\bigl(p_b^{\rm cs\,+},\vec p^{\rm \,\,cs}_{b\perp},\mu,\nu\bigr)\, S_{ij}(\vec p_{s\perp},\mu,\nu)\,  
\,,\\
\text{Regime 3:} \quad \frac{\df \si_3}{\df Q^2\,  \df p_T} &= \si_0 \sum_{i,j} H_{ij}(Q^2,\mu) 
\int\! \df^2 \vec p_{s\perp}\, \de\bigl(p_T - |\vec p_{s\perp} |\bigr)
\int\! \frac{\df \xi_a}{\xi_a}\, \frac{\df \xi_b}{\xi_b}\, \df (2 p^0_s)\, \de\Bigl[1 - \frac{\tau}{\xi_a \xi_b} - \frac{2 p^0_s}{Q} \Bigr]
\nn \\ & \quad \times
f_i(\xi_a,\mu)\, f_j(\xi_b, \mu)\, S_{ij}(2 p^0_s, \vec p_{s\perp},\mu)
\,,\end{align}
\end{widetext}
which we discuss in turn below. Here $\tau \equiv Q^2/s$, where $s$ is the \emph{hadronic} center-of-mass energy.
The predictions from these factorization theorems give the full cross section up to power corrections,
\begin{align}
\frac{\df \si}{\df Q^2  \df p_T} &= \frac{\df \si_1}{\df Q^2  \df p_T}\Bigl[1+
\ORd{\frac{p_T^2}{Q^2}}\Bigr]
\,,\nn \\
\frac{\df \si}{\df Q^2  \df p_T} &= \frac{\df \si_2}{\df Q^2  \df p_T}\Bigl[1+
\ORd{1-z,\frac{p_T^2}{(1-z)^2 Q^2}}\Bigr]
\,,\nn \\
\frac{\df \si}{\df Q^2  \df p_T} &= \frac{\df \si_3}{\df Q^2  \df p_T}\bigl[1+
\ord{1-z}\bigr]
\,.\end{align}
The intermediate regime involves the most expansions but allows for the independent resummation of logarithms, whereas in regime 1 and 3 the threshold parameter is constrained to be of a specific size. We discuss how to combine predictions from these regimes in \sec{combine}.

Regime 1 is described by the standard transverse momentum factorization. The hard function $H_{ij}$ characterizes the short-distance scattering of partons $i$ and $j$, where the sum over channels is restricted to $\{i,j\}= \{q,\bar q\}, \{\bar q, q\}, \{g,g\}$, since we consider color-singlet production. The transverse momentum dependent (TMD) soft function $S_{ij}$ encodes the contribution $\vec p_{s\perp}$ to the transverse momentum from soft radiation. 
The TMD beam function $B_i(x, \vec p_\perp,\mu,\nu)$ describes the extraction of the parton $i$ out of the proton with momentum fraction $x$ and transverse momentum $\vec p_\perp$. The transverse momentum due to perturbative initial-state radiation is described by matching the TMD beam function onto PDFs~\cite{Collins:1981uk,Collins:1984kg,Stewart:2009yx,Becher:2010tm,Collins:2011zzd,GarciaEchevarria:2011rb,Chiu:2012ir}\footnote{Often the TMD beam and soft function are combined into one object~\cite{Becher:2010tm,Collins:2011zzd,GarciaEchevarria:2011rb}. This is inconvenient here because regime 2 involves the TMD soft function with collinear-soft functions instead of the standard TMD beam functions. Though they are related, see \eq{consistency12}, they differ in the rapidity logarithms, see \eq{canonical}.}
\begin{align}\label{eq:B_to_f}
  B_i(x,\vec p_{\perp},\mu,\nu) &= \sum_{i'} \int\!\frac{\df \xi}{\xi}\, I_{ii'}\Bigl(\frac{x}{\xi}, \vec p_{\perp},\mu,\nu\Bigr) f_{i'}(\xi,\mu)
 \nn \\ & \quad \times
   \Big[1 + \ORd{\frac{\lqcd^2}{\vec p_\perp^{\,2}}}\Big]
\,,\end{align}
corresponding to the factorization of the two collinear modes in~\tab{modes1}.
The diagonal matching coefficients $I_{ii}$ contain threshold logarithms of $1-x/\xi$, which will be resummed in regime 2. The beam and soft functions have rapidity divergences, which we treat using the rapidity regulator of Refs.~\cite{Chiu:2011qc,Chiu:2012ir}. The resulting dependence on the rapidity renormalization scale $\nu$ will be used to sum the associated rapidity logarithms.\footnote{This resummation can also be achieved using the Collins-Soper equation~\cite{Collins:1984kg, deFlorian:2001zd, Collins:2011ca} or directly exponentiating the rapidity logarithms using consistency~\cite{Chiu:2007dg,Becher:2010tm}.}

Regime 3 is described by threshold factorization. The nonperturbative collinear modes combine into the (threshold) PDF. The $p_T$ measurement only probes the soft radiation, leading to a more differential soft function $S_{ij}(2p_s^0, \vec{p}_{s\perp},\mu)$. Here $p_s^0$ is the energy of the soft radiation, that arises from the threshold restriction,
\begin{align} \label{eq:mom_cons_3}
 Q^2 &= (\xi_a \Ecm - p_s^-)(\xi_b \Ecm - p_s^+)
 \nn \\ & 
 = \hat s - Q\, (p_s^-\,e^{-Y} + p_s^+\, e^Y) + \mathcal{O}[(1-z)^2Q^2]
 \nn \\
 &= Q^2\Bigl(\frac{\tau}{\xi_a \xi_b} + \frac{p_s^-\,e^{-Y}}{Q} + \frac{p_s^+\, e^Y}{Q} + \mathcal{O}[(1-z)^2]\Bigr)\,.  \end{align}
At hadronic threshold, where $1-\tau = 1-Q^2/s \ll 1$, $Y = \ord{1-z}$ and can be dropped. This implies that only the energy of the soft radiation is probed, $p_s^- + p_s^+ = 2 p_s^0$. In the next section, we will show that $Y$ can also be eliminated at partonic threshold. Note that the factorization theorem in this regime does not involve any rapidity divergences.

Regime 2 sits between 1 and 3.  The collinear-soft functions $\cS_{i}$ encodes the contribution from csoft radiation to the measurement. 
The $n$-collinear-soft function is defined as the following matrix element in \SCETp
\begin{align} \label{eq:CSdef}
\CS_i(p^-,\vec p_\perp) &=
\frac{1}{N_c}\,\langle 0 | {\rm Tr}\big[ {\bf{\overline{T}}} (X_n^\dagger(0) V_n(0))\, \de(p^- - {\bf P}^-)
\nn \\ & \quad \times
\de^2(\vec p_\perp - \vec {\bf P}_\perp) {\bf{T}} (V_n^\dagger(0) X_{n}(0))\big]| 0 \rangle \,.\end{align}
Here $X_n$ and $V_n$ are eikonal Wilson lines oriented along the $n$ and $\bn$ direction~\cite{Bauer:2011uc}, in the fundamental (adjoint) representation for $i=q, \bar q$ ($i=g)$. The operator $\bf P^\mu$ picks out the momentum of the collinear-soft radiation in the intermediate state, and $({\bf{\overline T}})$ ${\bf T}$ denotes (anti-)time ordering.
The matching onto the effective theory and decoupling of the modes follows from Refs.~\cite{Bauer:2011uc, Procura:2014cba}. An essential step in proving factorization involves the cancellation of Glauber gluons, which was shown in  Ref.~\cite{Laenen:2000ij} using the methods developed in Refs.~\cite{Bodwin:1984hc,Collins:1985ue,Collins:1988ig}. The convolution structure of the factorization theorem arises due to momentum conservation
\begin{align} \label{eq:mom_cons_2}
 Q^2 &= (\xi_a \Ecm - p^{\rm \,cs\, -}_{a})(\xi_b \Ecm - p^{\rm \,cs\, +}_{b})
 \nn \\ & 
 = \hat s - Q\, (p^{\rm \,cs\, -}_{a}\,e^{-Y} + p^{\rm \,cs\, +}_{b}\, e^Y) + \mathcal{O}[(1-z)^2Q^2]
 \,, \nn \\ 
  &= Q^2\Bigl(\frac{\tau}{\xi_a \xi_b} + \frac{p^{\rm \,cs\, -}_{a}\,e^{-Y}}{Q} + \frac{p^{\rm \,cs\, +}_{b}\, e^Y}{Q} + \mathcal{O}[(1-z)^2]\Bigr)\, ,\nn\\
  p_T &= |\vec p^{\rm \,cs}_{a\perp} + \vec p^{\rm \,cs}_{b\perp} + \vec p_{s\perp} |.
\end{align}
At hadronic threshold, $Y$ is again power suppressed and drops out. We will argue below why the same is true at partonic threshold.

\subsection{Partonic threshold}\label{sec:factor2}

If we can't eliminate $Y$ from \eq{mom_cons_3}, we would need a soft function that is differential in $p_s^-$, $p_s^+$ and $\vec{p}_{s\perp}$. However, boosting such a soft function leaves the Wilson lines invariant and changes the measurement to
\begin{align} \label{eq:soft_boost}
S_{ij}(p_s^- e^Y, p_s^+ e^{-Y},\vec{p}_{s\perp},\mu) = S_{ij}(p_s^-,p_s^+,\vec{p}_{s\perp},\mu)
\,.\end{align}
This implies that $Y$ can be eliminated form \eq{mom_cons_3} and the soft function only depends on the combination $2p_s^0 = p_s^- + p_s^+$ and $\vec{p}_{s\perp}$.

This argument does not immediately carry over to regime 2, since there the rapidity regulator breaks boost invariance,
\begin{align}
\cS_i\bigl(p_a^{\rm cs\,-} e^Y,\ \vec p^{\rm \,\,cs}_{a\perp},\mu,\nu\bigr) &=
    e^{-Y} \cS_i\bigl[p_a^{\rm cs\,-},\vec p^{\rm \,\,cs}_{a\perp},\mu,\nu/(e^{-Y})\bigr]
  \,,\nn \\
\cS_j\bigl(p_b^{\rm cs\,+} e^{-Y},\vec p^{\rm \,\,cs}_{b\perp},\mu,\nu\bigr) &=
    e^Y \cS_j\bigl[p_b^{\rm cs\,+}, \vec p^{\rm \,\,cs}_{b\perp},\mu,\nu/(e^{Y})\bigr]
\,.\end{align}
We can eliminate $Y$ using the rapidity evolution discussed in \sec{anom_2}, 
\begin{align}
\cS_i\bigl[p_a^{\rm cs\,-},\vec p^{\rm \,\,cs}_{a\perp},\mu,\nu/(e^{-Y})\bigr] &= 
\int\! \df^2 \vec p_{a\perp}^{\rm \,\,cs\,'} U^i_\nu(\vec p^{\rm \,\,cs\,'}_{a\perp},\mu,\nu e^Y, \nu)
\nn \\ & \quad \times
 \cS_i\bigl[p_a^{\rm cs\,-},\vec p^{\rm \,\,cs}_{a\perp} \!-\! \vec p^{\rm \,\,cs\,'}_{a\perp},\mu,\nu\bigr]
\,,\nn \\
\cS_j\bigl[p_b^{\rm cs\,+},\vec p^{\rm \,\,cs}_{b\perp},\mu,\nu/(e^Y)\bigr] &= 
\int\! \df^2 \vec p_{b\perp}^{\rm \,\,cs\,'}\, U^j_\nu(\vec p^{\rm \,\,cs\,'}_{b\perp},\mu, \nu e^{-Y},\nu)
\nn \\ & \quad \times
 \cS_j\bigl[p_b^{\rm cs\,+},\vec p^{\rm \,\,cs}_{b\perp} \!-\! \vec p^{\rm \,\,cs\,'}_{b\perp},\mu,\nu\bigr]
\,.\end{align}
The evolution kernels cancel against each other in the final result 
\begin{align}
&\int\! \df^2 \vec p^{\rm \,\,cs}_{a\perp}\,\df^2 \vec p^{\rm \,\,cs}_{b\perp}\,
\de\bigl(p_T - |\vec p^{\rm \,\,cs}_{a\perp} + \vec p^{\rm \,\,cs}_{b\perp} + \vec p_{s\perp} |\bigr)
\nn \\ & \quad \times
U^i_\nu(\vec p^{\rm \,\,cs}_{a\perp}- \vec p^{\rm \,\,cs \,'}_{a\perp},\mu,\nu e^Y, \nu)\, 
U^i_\nu(\vec p^{\rm \,\,cs}_{b\perp}- \vec p^{\rm \,\,cs \,'}_{b\perp},\mu,\nu e^{-Y},\nu)\, 
\nn \\ & \quad
= \de\bigl(p_T - |\vec p^{\rm \,\,cs \,'}_{a\perp} + \vec p^{\rm \,\,cs \,'}_{b\perp} + \vec p_{s\perp} |\bigr)
\,,\end{align}
so $Y$ may also be dropped from \eq{mom_cons_2} at partonic threshold.

\subsection{Consistency relations}\label{sec:factor3}

In the threshold limit, the factorization theorem for regime 1 should match onto regime 2. This leads to the following consistency relation for the fixed-order content
\begin{align}\label{eq:consistency12}
I_{ij}(y, \vec p_{\perp},\mu,\nu)
& = \de_{ij}\, Q\,\cS_i[(1-y)Q ,\vec p_{\perp},\mu,\nu] 
\nn \\ & \quad \times
[1+ \ord{1-y}]
\,,\end{align}
where $y=x/\xi$.
Similarly, consistency of the factorization theorems in regimes 3 and 2 implies that
\begin{align}\label{eq:consistency23}
&S_{ij}(2p^0, \vec p_\perp, \mu)
 \\
& \quad
 = \int\! \df^2 \vec p^{\rm \,\,cs}_{a\perp}\,\df^2 \vec p^{\rm \,\,cs}_{b\perp}\,\df^2 \vec p_{s\perp}\,  
\de\bigl[\vec p_\perp - (\vec p^{\rm \,\,cs}_{a\perp} + \vec p^{\rm \,\,cs}_{b\perp} + \vec p_{s\perp}) \bigr]
\nn   \\ & \qquad \times
\int\! \df p_a^{\rm cs\,-}\, \df p_b^{\rm cs\,+}\,\de\bigl[2p^0 - \bigl(p_a^{\rm cs\,-} + p_b^{\rm cs\,+}\bigr) \bigr]\,  
\nn \\ & \qquad \times
\cS_i\bigl(p_a^{\rm cs\,-},\vec p^{\rm \,\,cs}_{a\perp},\mu,\nu\bigr)\,
\cS_j\bigl(p_b^{\rm cs\,+},\vec p^{\rm \,\,cs}_{b\perp},\mu,\nu\bigr)\,
\nn   \\ & \qquad \times
S_{ij}(\vec p_{s\perp},\mu,\nu)\,\Bigl[1+ {\cal O}\Bigl({\frac{p_T^2}{p_0^2}}\Bigr)\Bigr]
\,.\nn\end{align}
where $2p^0 = (1-z)Q$.
Note that the rapidity divergences must cancel between the csoft functions and the TMD soft function on the right-hand side, since the double differential soft function does not have them.
We verify these consistency equations at NLO in \sec{1loop}.

\section{One-loop ingredients}\label{sec:1loop}

In this section we give the one-loop soft and csoft functions. We verify the consistency relations in \eqs{consistency12}{consistency23} between the different regimes, using these expressions.

\subsection{Soft function}

For completeness, we start by giving the one-loop TMD soft function~\cite{Chiu:2012ir}
\begin{align} \label{eq:TMDsoft}
S^{(1)}_{ij}(\vec p_\perp,\mu,\nu) &=  \frac{\alpha_s C_i}{\pi^2} \biggl[ 
- \frac{1}{\mu^2} \cL_1 \Big( \frac{p_T^{\,2}}{\mu^2}\Big)
+ \frac{2}{\mu^2} \cL_0 \Big( \frac{p_T^{\,2}}{\mu^2}\Big) \ln \frac{\nu}{\mu}
\nn \\ & \quad  
-\frac{\pi^2}{12}\, \delta(p_T^{\,2}) \biggr]
\,.\end{align}
Here $p_T^2 = -p_\perp^{\,2}$, the color factor $C_i$ is $C_F$ for quarks and $C_A$ for gluons, and the plus distributions are defined as
\begin{align} \label{eq:plusdef}
\cL_n(x)
&\equiv \biggl[ \frac{\theta(x) \ln^n x}{x}\biggr]_+
\nn \\
& = \lim_{\beta \to 0} \biggl[
  \frac{\theta(x- \beta)\ln^n x}{x} +
  \delta(x- \beta) \, \frac{\ln^{n+1}\!\beta}{n+1} \biggr]
\,.\end{align} 

Using the approach in Ref.~\cite{Kasemets:2015uus}, we obtain the double differential soft function at one-loop order\footnote{Azimuthal symmetry implies $\de^{(2)}(\vec p_\perp - \dots) = \de(p_T^2 - \dots) / \pi$, allowing us to eliminate vector quantities.}
\begin{align} \label{eq:S_p0_pT}
S_{ij}^{(1)} (2p^0, \vec p_\perp ,\mu) &= \frac{\al_s C_i}{\pi^2}  \frac{\df}{\df (2p^0)}\frac{\df}{\df (p_T^2)}
\bigg\{
\theta(p^0) \theta(p_T) 
 \\ & \quad \times
\bigg[  2 \ln^2 \frac{2p^0}{\mu} 
- \frac{\pi^2}{12}
+ \theta(p^0 - p_T) 
\nn \\ & \quad \times
\Big( 2 a^2
- 4a \ln \frac{2p^0}{p_T}  
+ 2 {\rm Li}_2 \big(-e^{-2a}\big)\Big)
\bigg]
\bigg\}
\,,\nn\end{align}
where $a= {\rm arccosh} (p_0/p_T)$. This is directly related to the fully-differential soft function of Ref.~\cite{Mantry:2009qz}. The projection from $p^+$ and $p^-$ onto $2p^0$ does not affect the renormalization but is responsible for the complicated finite terms above.

\subsection{Collinear-soft function}
\label{sec:csoft}

At first sight, the csoft functions in \eq{CSdef} appear identical to those for the joint resummation of transverse momentum and the beam thrust event shape in Ref.~\cite{Procura:2014cba}. They involve the other light-cone component, but the calculation is symmetric under $p^- \lra p^+$. However, the zero-bin~\cite{Manohar:2006nz} that accounts for the overlap with other modes differs. In~\citere{Procura:2014cba} the zero-bin vanished in pure dimensional regularization, converting all IR divergences into UV divergences. Here, the zero-bin that accounts for the overlap with collinear radiation with energy $(1-z)Q$ plays a similar role, but there is also a non-trivial zero-bin from the overlap with soft radiation.\footnote{Here we find it convenient to not expand the rapidity regulator $|k^+ - k^-|^{-\eta}$ of \citere{Chiu:2011qc,Chiu:2012ir} according to the power counting of each mode. This distinction is irrelevant for the soft function where $k^+$ and $k^-$ are of the same parametric size. Thus the same is true for all ingredients, by exploiting consistency of the various factorization theorems. When expanding the regulator  $|k^+ - k^-|^{-\eta} \to |k^-|^{-\eta}$, the zero-bin is scaleless. However, the regulator now explicitly breaks the $p^- \lra p^+$ symmetry, so the collinear-soft function is not the same as in \citere{Procura:2014cba}.} This leads to
\begin{align} \label{eq:zero-bin}
\cS_i \bigl(p^-,\vec p_\perp,\mu,\nu\bigr) &= \int \df^2 \vec p^{\,'}_\perp\, \cS_i^\text{Ref.\cite{Procura:2014cba}} \bigl(p^-,\vec p_\perp- \vec p^{\,'}_\perp,\mu,\nu\bigr)
\nn \\ & \quad \times
S_{i\bar i}^{-1} \bigl(\vec p^{\,'}_\perp,\mu,\nu\bigr)
\,.\end{align}
Here the collinear-soft function of \citere{Procura:2014cba} is taken, which is a function of $p^+$ and $p_\perp$, and $p^+$ is evaluated at $p^-$. $S_{i\bar i}^{-1}$ is the inverse of the TMD soft function. 
(The relation between zero-bins and inverse soft functions has been discussed in e.g.~Refs.~\cite{Lee:2006nr,Idilbi:2007ff}.)

From the expression in~\citere{Procura:2014cba} and the TMD soft function in \eq{TMDsoft}, we obtain the one-loop csoft function
\begin{align} \label{eq:csoft}
\cS^{(1)}(p^-,\vec p_\perp,\mu,\nu) &=   \frac{\alpha_s C_i}{\pi^2} \biggl[
 \frac{1}{\mu} \cL_0 \Big(\frac{p^-}{\mu}\Big) \frac{1}{\mu^2} \cL_0 \Big(\frac{p_T^2}{\mu^2}\Big)
\nn \\ & \quad  
-  \frac{1}{\mu^2} \cL_0 \Big(\frac{p_T^2}{\mu^2}\Big) \ln \frac{\nu}{\mu} \de(p^-)
\biggr]
\,.\end{align}

\subsection{Consistency of the NLO ingredients}

We have verified the consistency of regimes 1 and 2, as expressed in \eq{consistency12}. The expressions for the TMD beam functions with the $\nu$ rapidity regulator are given at one-loop order in Refs.~\cite{Chiu:2012ir, Ritzmann:2014mka}. They can directly be compared to the one-loop csoft function in \eq{csoft}. 

At one-loop order the consistency relation between regime 2 and 3 in \eq{consistency23} reads
\begin{align}\label{eq:consistency231L}
S^{(1)}_{ij} \bigl(2p^0, \vec p_\perp, \mu \bigr) 
&=  \bigl[2\cS^{(1)}_i\bigl(2p^0,\vec p_\perp,\mu,\nu\bigr) 
 \\ & \quad
 + \de(2p^0)\,S^{(1)}_{ij}(\vec p_\perp,\mu,\nu)\bigr]
 \Big[1+ {\cal O}\Big({\frac{p_T^2}{p_0^2}}\Big)\Big]
\,.\nn\end{align}
Expanding the double-differential soft function in \eq{S_p0_pT},
\begin{align} 
S^{(1)}_{ij} (2p^0,\vec p_\perp,\mu) &
= \frac{\al_s C_i}{\pi^2} \bigg[ 
 - \frac{1}{\mu^2} \cL_1 \Big( \frac{p_T^{\,2}}{\mu^2}\Big) \de(2p^0)
  \\ & \quad
 + \frac{2}{\mu} \cL_0 \Big(\frac{2p_0}{\mu}\Big)\, \frac{1}{\mu^2} \cL_0 \Big( \frac{p_T^{\,2}}{\mu^2} \Big)
\nn \\ & \quad
- \frac{\pi^2}{12} \de(2p^0) \delta(p_T^2) 
\bigg]\,\Big[1+ {\cal O}\Big({\frac{p_T^2}{p_0^2}}\Big)\Big]
\,,\nn\end{align}
it is straightforward to verify this using the expression for the TMD soft function in \eq{TMDsoft} and the csoft function in \eq{csoft}. 

\section{Anomalous dimensions and consistency}
\label{sec:anom}

In this section we collect the renormalization group (RG) equations for the ingredients of the factorization theorems, which are needed for resummation.
We also verify the consistency of the anomalous dimensions. 

\subsection{Regime 1}

We start by considering the ingredients that enter the factorization theorem for regime 1. The anomalous dimension of the hard function is
\begin{align}
&\mu\frac{\text{d}}{\text{d}\mu}H_{ij}(Q^2,\mu) = \gamma_{H}^i(Q^2,\mu)H_{ij}(Q^2,\mu)
\,,\nn \\ 
&\gamma_{H}^i(Q^2,\mu) = 2\Gamma_{\text{cusp}}^i(\al_s) \ln\Bigl(\frac{Q^2}{\mu^2}\Bigr) + \gamma_H^i(\alpha_s)\,.
\end{align}
Here $\Gamma_{\text{cusp}}^i(\al_s)$ is the cusp anomalous dimension~\cite{Korchemsky:1987wg, Moch:2004pa} and $\gamma_H^i(\alpha_s)$ the non-cusp term~\cite{Kramer:1986sg,Matsuura:1988sm,Harlander:2000mg,Moch:2005id,Idilbi:2005ni,Moch:2005tm,Idilbi:2006dg}.

The renormalization of the TMD beam functions has the following structure%
\footnote{Unlike $\mu$-anomalous dimensions, the structure of the $\nu$-anomalous dimension changes at each order. Thus the structure of $\gamma_\nu$ shown here is only valid at one-loop order. The higher-order expressions follows from $\df \gamma_\nu/ \df \ln \mu = \df \gamma_\mu / \df \ln \nu$~\cite{Chiu:2012ir,Luebbert:2016itl}.}
\begin{align} \label{eq:ga_B}
&\mu\frac{\text{d}}{\text{d}\mu}\,B_i(x,\vec{p}_{\perp},\mu,\nu) = \gamma^i_B\Bigl(\mu,\frac{\nu}{\w}\Bigr)\,B_i(x,\vec{p}_{\perp},\mu,\nu)
\,, \\
&\,\nu\frac{\text{d}}{\text{d}\nu}\,B_i(x,\vec{p}_{\perp},\mu,\nu) = 
\int \df^2 \vec p_{\perp}^{\,'} \gamma_\nu^i (\vec p\,_{\perp} - \vec p_{\perp}^{\,'},\mu)
\nn \\ & 
\phantom{\nu\frac{\text{d}}{\text{d}\nu}B_i(x,\vec{p}_{\perp},\mu,\nu) =} \times
B_i(x,\vec{p}_{\perp}^{\,'},\mu,\nu)
\,,\nn \\ &  
\gamma_B^i\Bigl(\mu,\frac{\nu}{\w}\Bigr) = 2\Gamma_{\text{cusp}}^i (\al_s) \ln\Big(\frac{\nu}{\w}\Big) + \gamma_B^i(\alpha_s)
\,,\nn \\ 
& \hspace{0.8ex} \gamma_\nu^i(\vec p_\perp,\mu) = - \Ga_{\text{cusp}}^i(\al_s)\, \frac{1}{\pi}\, \frac{1}{\mu^2}\, \cL_0\Bigl(\frac{p_T^2}{\mu^2}\Bigr) + \gamma_\nu^i(\al_s) \de^{(2)} (\vec{p}_{\perp})
\,,\nn\end{align}
where $\w = x \sqrt{s}$ characterizes its rapidity. 
The non-cusp anomalous dimension $\ga_B^i(\al_s)$ has been calculated to two-loop order~\cite{Luebbert:2016itl}, and the rapidity anomalous dimension $\ga_\nu$ was recently determined at three loops~\cite{Li:2016ctv}.

The TMD soft function has the anomalous dimension
\begin{align}
& \mu\frac{\text{d}}{\text{d}\mu}\,S_{ij}(\vec{p}_{\perp},\mu,\nu) = \gamma_{S}^{i}(\mu,\nu)S_{ij}(\vec{p}_{s\perp},\mu,\nu)
\,,\nn \\ 
& \nu\frac{\text{d}}{\text{d}\nu}\,S_{ij}(\vec{p}_{\perp},\mu,\nu) = -2\! \int \df^2 \vec p\,'_{\perp} \gamma_\nu^i (\vec p\,_{\perp} \!-\! \vec p\,'_{\perp},\mu)S_{ij}(\vec{p}\,'_{\perp},\mu,\nu)
\,,\nn \\
& \,\,\,\gamma_{S}^i(\mu,\nu) = 4\Gamma_{\text{cusp}}^i(\al_s) \ln\Bigl(\frac{\mu}{\nu}\Bigr) + \gamma_S^i(\alpha_s)
\,,\end{align}
where the non-cusp anomalous dimension $\ga_S^i(\alpha_s)$ is known to two-loop order~\cite{Luebbert:2016itl}.
We wrote its $\nu$-anomalous dimension in terms of that of the TMD beam function, exploiting consistency of the factorization theorem in regime 1. The $\mu$-independence of the cross section implies the following consistency relation
\begin{align}
\gamma_{H}^i(Q^2,\mu) + \gamma_B^i\Bigl(\mu,\frac{\nu}{\w_a}\Bigr) + \gamma_B^j\Bigl(\mu,\frac{\nu}{\w_b}\Bigr) + \gamma_{S}^i(\mu,\nu)  = 0
\,,\end{align}
which is straightforward to verify using $Q^2 = \w_a \w_b$.

\subsection{Regime 2}
\label{sec:anom_2}

We have two new ingredients in regime 2, the PDF and the collinear-soft function. 
In the threshold limit, the mixing between PDFs of different flavors is suppressed and the anomalous dimension simplifies to~\cite{Korchemsky:1992xv}
\begin{align}
&\mu\frac{\text{d}}{\text{d}\mu}\,f_i(\xi,\mu) =\int_{\xi}^1\! \df\xi'\, \gamma_f^i(\xi'-\xi,\mu)\, f_i(\xi',\mu)
\\
&\gamma_f^{i}(y,\mu) = 2\Gamma_{\text{cusp}}^i(\al_s)\, \cL_0(y) + \gamma_f^i(\alpha_s)\, \de(y)\,.
\nn \end{align}
The anomalous dimensions of the PDF have been calculated up to three loops~\cite{Vogt:2004mw, Moch:2004pa}.

The collinear-soft function has the following anomalous dimension
\begin{align}
& \mu\frac{\text{d}}{\text{d}\mu}\mathscr{S}_i(p^-,\vec{p}_{\perp},\mu,\nu) = 
\int \text{d} p'^{-}\,\gamma_{\mathscr{S}}^i\Bigl(\frac{p^--p'^-}{\nu},\mu \Bigr)
\nn \\ & 
\phantom{ \mu\frac{\text{d}}{\text{d}\mu}\mathscr{S}_i(p^-,\vec{p}_{\perp},\mu,\nu) = } \times \mathscr{S}_i(p'^-,\vec{p}_{\perp},\mu,\nu)
\,, \nn \\ 
& \nu\frac{\text{d}}{\text{d}\nu}\mathscr{S}_i(p^-,\vec{p}_{\perp},\mu,\nu) = \int \df^2 \vec p\,'_{\perp}\, \gamma_\nu^i (\vec p\,_{\perp} - \vec p\,'_{\perp},\mu)
\nn \\ & 
\phantom{\nu\frac{\text{d}}{\text{d}\nu}\mathscr{S}_i(p^-,\vec{p}_{\perp},\mu,\nu) =} \times
\mathscr{S}_i(p^-,\vec p\,'_{\perp},\mu,\nu)
\,, \\ &
\gamma_{\mathscr{S}}^i\Bigl(\frac{p^-}{\nu},\mu \Bigr) = -2 \Ga_{\text{cusp}}^i(\al_s)\, \frac{1}{\nu}\, \cL_0\Bigl(\frac{p^-}{\nu}\Bigr) + \gamma_{\mathscr{S}}^i(\al_s)\de (p^-)\nn
\,.\end{align}
We exploited consistency to write the rapidity anomalous dimension in terms of $\ga_\nu$ in \eq{ga_B}, which agrees with our one-loop calculation in \eq{csoft}. 
The consistency relation between the $\mu$-anomalous dimensions reads 
\begin{align} \label{eq:mu_cons_2}
\big[\gamma_H^i(Q^2,\mu)+\gamma_S^i(\mu,\nu)\big]\delta(1-\xi) + 2\gamma_f^{i}(\xi,\mu) &
\nn \\ 
+ 2Q\,\gamma_{\mathscr{S}}^i\Bigl[\frac{(1-\xi)Q}{\nu},\mu \Bigr] &= 0
\,.\end{align}
\eq{mu_cons_2} implies for the non-cusp anomalous dimension
\begin{align}
\gamma_{\mathscr{S}}^i(\al_s) = - \frac12 \bigl[\ga_H^i(\al_s) + 2\ga_f^i (\al_s) + \ga_S^i(\al_s)\bigr]
\,,\end{align}
which vanishes up to two-loop order. Alternatively, the zero-bin in \eq{zero-bin} and consistency of the \SCETp factorization in \citere{Procura:2014cba} imply that
\begin{align}
\gamma_{\mathscr{S}}^i(\al_s) &= - \frac12 \bigl[\ga_H^i(\al_s) + 2\ga_B^i (\al_s) + \bar \ga_{S}^i(\al_s)\bigr] - \ga_S^i(\al_s)
\nn \\ &
= - \frac12 \bigl[\ga_S^i (\al_s) + \bar \ga_{S}^i(\al_s)\bigr] 
\,,\end{align}
where $\bar \ga_{S}^i(\al_s)$ is the non-cusp anomalous dimension for the (beam)thrust soft function. We have verified this at two-loop order.

\subsection{Regime 3}

For regime 3 we need the anomalous dimension of the double differential soft function, 
\begin{align}
& \mu\frac{\text{d}}{\text{d}\mu}S_{ij}(2p^0,\vec{p}_{\perp},\mu) = 
\int \text{d} (2p'^0)\,\tilde{\gamma}_S^i (2p^0-2p'^0,\mu)
 \\ & 
\phantom{ \mu\frac{\text{d}}{\text{d}\mu}S_{ij}(2p^0,\vec{p}_{\perp},\mu) =}
\times S_{ij}(2p'^0,\vec{p}_{\perp},\mu)
\,, \nn \\ 
&\tilde{\gamma}_S^i(2p^0,\mu) = -4 \Ga_{\text{cusp}}^i(\al_s)\, \frac{1}{\mu}\, \cL_0\Bigl(\frac{2p^0}{\mu}\Bigr) + \tilde{\gamma}_S^i(\al_s)\, \de(2p_0)
\,. \nn\end{align}
Here we included a tilde on $\ga_S$ to distinguish it from the anomalous dimension of the TMD soft function.
Consistency of the factorization theorem in regime 3 implies that the anomalous dimensions satisfy
\begin{align} \label{eq:thres_consistency}
\gamma_H^i(Q^2,\mu)\delta(2p^0) + \frac{2}{Q}\gamma_f^{i}\Bigl(\frac{Q-2p^0}{Q},\mu\Bigr)& 
 \\
+ \tilde{\gamma}_S^i(2p^0 ,\mu) &= 0
\,. \nn\end{align}
This implies that the anomalous dimension $\tilde{\gamma}_S$ is equal to that of the threshold soft function in Ref.~\cite{Becher:2007ty}. It also implies the following all-orders relationship between the threshold and (beam)thrust soft function
\begin{align}
   \tilde{\gamma}_S^i(\al_s) =  -\bar{\gamma}_S^i(\al_s) 
\,.\end{align}
This result also follows from the consistency relation for DIS in the threshold limit~\cite{Becher:2006mr}
\begin{align}
  \ga_H^i + \ga_J^i + \ga_f^i = 0
\,,\end{align}
where $\ga_J^i$ is the non-cusp anomalous dimension of the jet function,  together with the consistency of threshold (\eq{thres_consistency}) and beam thrust factorization for Drell-Yan~\cite{Stewart:2009yx}.

\section{Resummation}\label{sec:resummation}

We now discuss how to achieve the resummation using the RG evolution. We identify the natural scales, and explicitly show how to include the RG evolution in the factorization theorem for regime 1. A procedure to combine the resummed predictions from the different regimes is also described.

\subsection{Scales and evolution}

From the anomalous dimensions in \sec{anom}, we can immediately read off the natural scales for the perturbative ingredients
  \begin{align} \label{eq:canonical}
    \mu_H &\sim  Q
    \,, \nn \\    
    \mu_B &\sim p_T\,, \quad \nu_B \sim \w \sim Q
    \,, \nn \\
    \mu_\cS &\sim p_T \,, \quad \nu_\cS \sim (1-z) Q
    \,, \nn \\
    \mu_S &\sim p_T\,, \quad \nu_S \sim p_T 
  \,.\end{align}
The resummation of logarithms of $p_T/Q$ and $1-z$ is achieved by evaluating each ingredient at its natural scale, where it contains no large logarithms, and evolving them to a common $\mu$ and $\nu$. The ingredients needed at various orders in resummed perturbation theory are summarized in Table~\ref{tab:ingredients}. 
\begin{table}[h]
\centering
   \begin{tabular}{c|cccc} 
     \hline \hline
     Order & $H, B, S, \cS, f$ &  $\gamma_{X}^i$ & $\Ga_{\text{cusp}}^i$ & $\beta$\\
     \hline
     LL & LO & & \,1-loop\,& \,1-loop\, \\
     NLL & LO & \,1-loop\,& \,2-loop\,& \,2-loop\, \\
     NNLL & NLO & 2-loop & 3-loop & 3-loop\\
     NNNLL & NNLO & 3-loop & 4-loop & 4-loop\\
     \hline \hline
   \end{tabular} 
   \caption{
   Ingredients required at each order in resummed perturbation theory. The columns correspond to the fixed-order content, non-cusp ($X=H,B,S,\cS,f,\nu$) and cusp anomalous dimension, and the QCD beta function.}
    \label{tab:ingredients}
\end{table}
To illustrate how to achieve this resummation in the cross section, we show explicitly how to include the evolution kernels for regime 1,
\begin{widetext}
\begin{align}
\frac{\df \si_1}{\df Q^2\, \df p_T} &= \si_0 \sum_{i,j} H_{ij}(Q^2,\mu_H) 
\int\! \df^2 \vec p_{a\perp}\,\df^2 \vec p_{b\perp}\,\df^2 \vec p_{s\perp}\, \de\bigl(p_T - |\vec p_{a\perp} + \vec p_{b\perp} + \vec p_{s\perp} |\bigr)
\int\! \df x_a\, \df x_b\, \de(\tau - x_a x_b)\,
\nn  \\ & \quad \times
\int\! \df^2 \vec p_{a\perp}^{\,'}\,\df^2 \vec p_{b\perp}^{\,'}\,\df^2 \vec p_{s\perp}^{\,'}\, B_i(x_a, \vec p_{a\perp} - \vec p_{a\perp}^{\,'},\mu_B,\nu_B)\,B_j(x_b,\vec p_{b\perp} - \vec p_{b\perp}^{\,'},\mu_B,\nu_B)\, S_{ij}(\vec p_{s\perp} - \vec p_{s\perp}^{\,'}, \mu_S, \nu_S)
\nn  \\ & \quad \times
U_\nu^i(\vec p_{a\perp}^{\,'},\mu_B,\nu,\nu_B)\,U_\nu^j(\vec p_{b\perp}^{\,'},\mu_B,\nu,\nu_B)
\int\! \df^2 \vec k_\perp\, U_\nu^i(\vec p_{s\perp}^{\,'} - \vec k_\perp, \mu_S,\nu_S,\nu) U_\nu^j(\vec k_\perp,\mu_S,\nu_S,\nu)
\nn  \\ & \quad \times
\exp\bigg[ \int_{\mu_H}^{\mu}\frac{\text{d}\mu'}{\mu'}\gamma_H^i(Q^2,\mu') +
\int_{\mu_B}^{\mu}\frac{\text{d}\mu'}{\mu'} 2 \gamma_B^i(\mu',\nu) +
 \int_{\mu_S}^{\mu}\frac{\text{d}\mu'}{\mu'}\gamma_S^i(\mu',\nu)\bigg]
\,. \end{align}
\end{widetext}
The rapidity evolution kernel $U_\nu$ of the beam function is defined through
\begin{align}
&\nu\frac{\text{d}}{\text{d}\nu}\,U_\nu^i(\vec{p}_{\perp},\mu,\nu,\nu_0) = 
\int \df^2 \vec p_{\perp}^{\,'} \gamma_\nu^i (\vec p\,_{\perp} - \vec p_{\perp}^{\,'},\mu)
\nn \\ & 
\phantom{\nu\frac{\text{d}}{\text{d}\nu}\,U_\nu^i(\vec{p}_{\perp},\mu,\nu,\nu_0) = } \times
U_\nu^i(\vec{p}_{\perp}^{\,'},\mu,\nu,\nu_0)
\,,\nn \\ & 
U_\nu^i(\vec{p}_{\perp},\mu,\nu_0,\nu_0) = \de^{(2)}(\vec{p}_{\perp})
\,.\end{align}
We write the rapidity evolution of the soft function in terms of this, exploiting  that its rapidity anomalous dimension differs by a factor of -2. In the next section we will argue that we can obtain the cross section in regime 2 from the one in regime 1 by adjusting the scale choice. The resummation in regime 3 has a different structure.

\subsection{Combining regimes}
\label{sec:combine}

The matching relation in \eq{consistency12} and the scales in \eq{canonical} imply that simply choosing
\begin{align}
 \nu_B \sim (1-z) Q\,,
\end{align}
smoothly interpolates between regime 1 and 2,
\begin{align}
  \frac{\df \si_{1+2}}{\df Q^2\, \df p_T} = \frac{\df \si_1}{\df Q^2\, \df p_T}\biggr|_{\nu_B \sim (1-z) Q}\,.
\end{align}
Ref.~\cite{Procura:2011aq} noted that such a scale choice removes the large logarithms in the anomalous dimension of the beam function coefficient $I_{ii}$, since
\begin{align}
 \ga_{I_{ii}}\Bigl(y,\mu,\frac{\nu}{\w}\Bigr) &= \ga_B^i\Bigl(\mu,\frac{\nu}{\w}\Bigr)\, \de(1-y) - \ga_f^i(y,\mu)
\nn \\ &
 =\biggl[2 \,\Ga^i_\text{cusp}(\al_s) \Bigl(\ln \frac{\nu}{\w}\, \de(1-y) 
 -  \cL_0(1-y) \Bigr)
  \nn \\ & \quad 
  + \bigl(\ga_B^i(\al_s) - \ga_f^i(\al_s)\bigr) \de(1-y)\biggr]
  \nn \\ & \quad \times
   [1+ \ord{1-y}]
\,,\end{align}
in the threshold limit. The factorization analysis we perform here establishes that this indeed sums \emph{all} threshold logarithms in regime 2.

As an aside, we note that this implies that the conjecture of Ref.~\cite{Echevarria:2016scs} is correct. There it was stated that for the beam function in the threshold limit the coefficient of the $\cL_0(1-y)$ term in the matching coefficient $\cI_{ii}(y,\vec p_\perp,\mu,\nu)$ is the rapidity anomalous dimension $-\ga_{\nu}^i$. The conjecture was formulated in impact parameter space $b_\perp$ (and requires modification for $p_\perp$). Its validity follows from our framework, since \eq{consistency12} relates it to the corresponding term in the csoft function, whose nontrivial $\nu_\cS \sim (1-y)Q$ and $\mu_\cS \sim p_\perp$ dependence is fully generated by the $\nu$ and $\mu$ RGE, respectively.\footnote{In the beam function it is not a priori clear that all logarithms of $1-y$ are generated by the $\nu$ RGE, because $1-y$ does not have an (independent) power counting associated with it.}

We now discuss how to combine regimes 1 and 2 with 3, which involves a nontrivial matching. 
Implementing this additively,
\begin{align}
  \frac{\df \si_{1+2+3}}{\df Q^2\, \df p_T} &= \frac{\df \si_2}{\df Q^2\, \df p_T}
   + \biggl(\frac{\df \si_1}{\df Q^2\, \df p_T} - \frac{\df \si_2}{\df Q^2\, \df p_T}\biggr|_{\nu_\cS = \nu_B} \biggr)
  \nn \\ & \quad
   + \biggl(\frac{\df \si_3}{\df Q^2\, \df p_T} - \frac{\df \si_2}{\df Q^2\, \df p_T}\biggr|_{\nu_\cS = \nu_S} \biggr)
\,,\end{align}
where for example the subscript $\nu_\cS = \nu_B$ indicates that the additional threshold resummation of regime 2 has been turned off in this term. Note that regime 2 plays a crucial role to account for the overlap between regimes 1 and 3. To smoothly turn off the resummation as one approaches regimes 1 and 3, requires the use of profile functions~\cite{Ligeti:2008ac, Abbate:2010xh}. 

The fixed-order QCD cross section $\si_{\rm FO}$ contains additional non-logarithmic corrections not contained in $\si_{1+2+3}$. They can be included in a similar manner,
\begin{align}
  \frac{\df \si}{\df Q^2\, \df p_T} &= \frac{\df \si_{1+2+3}}{\df Q^2\, \df p_T}
   + \biggl(\frac{\df \si_{\rm FO}}{\df Q^2\, \df p_T} - \frac{\df \si_{1+2+3}}{\df Q^2\, \df p_T}\biggr|_{\mu_i  = \nu_i = Q} \biggr)
   \nn \\ 
&= \frac{\df \si_{1+2+3}}{\df Q^2\, \df p_T}
   + \biggl(\frac{\df \si_{\rm FO}}{\df Q^2\, \df p_T} - \frac{\df \si_1}{\df Q^2\, \df p_T}\biggr|_{\mu_i  = \nu_i = Q} 
    \nn \\ & \quad
   - \frac{\df \si_{3}}{\df Q^2\, \df p_T}\biggr|_{\mu_i  = Q}
   + \frac{\df \si_2}{\df Q^2\, \df p_T}\biggr|_{\mu_i  = \nu_i = Q} \biggr)
\,.\end{align}

\section{Conclusions}\label{sec:conclusions}

In this letter, we developed a framework for the joint resummation of threshold and transverse momentum logarithms using SCET. There are three kinematic regimes, each with their own modes and all-orders factorization theorems. We discussed how these can be used to obtain resummed predictions, and how to combine the descriptions of the different regimes. Regime 2 is directly related to regime 1 through a change of scale choice, but regime 3 provides nontrivial corrections starting at NNLL. We checked the consistency of the individual factorization theorems from anomalous dimensions, as well as the consistency between different regimes. We also provided all ingredients necessary for NNLL resummation. In fact, all ingredients for NNNLL resummation can now be obtained from the literature, apart from the four-loop cusp anomalous dimension.\footnote{The three-loop non-cusp $\mu$-anomalous dimension for the TMD beam and soft function are not known either, but since they have the same natural $\mu$ scale this does not affect the central value but only the uncertainty of predictions. These anomalous dimensions depend on the scheme used to treat the rapidity divergences.}
The two-loop TMD beam and soft function were calculated in Ref.~\cite{Catani:2011kr,Catani:2012qa,Gehrmann:2012ze,Gehrmann:2014yya,Echevarria:2015byo,Luebbert:2016itl,Echevarria:2016scs} and the two-loop double differential soft function can be extracted from Ref.~\cite{Li:2011zp}. 
We note that this same approach can be used to describe heavy particle production in the presence of a veto on jets with $p_T^\text{jet} > p_T^\text{cut}$, where instead of transverse momentum logarithms the cross section contains logarithms of $p_T^\text{cut}/Q$. The convolutions in $p_T$ are replaced by multiplications where each ingredient depends on $p_T^\text{cut}$, but the framework is otherwise the same.

\acknowledgements

We thank D.~Neill, E.~Laenen, P.~Pietrulewicz and M.~Procura for discussions and comments on the manuscript.
This work was supported by the Netherlands Organization for Scientific Research (NWO) through a VENI grant (project number 680-47-448), and the D-ITP consortium, a program of the NWO that is funded by the Dutch Ministry of Education, Culture and Science (OCW). \\

\noindent {\bf Note Added:} \\
While this manuscript was in preparation \citere{Li:2016axz} appeared, which identified the same regimes and factorization theorems in position space.
Their focus was on using the threshold restriction as a rapidity regulator to simplify the calculation of the TMD soft function, see also \citere{Li:2016ctv}.
Instead we focus on deriving a framework for joint transverse momentum and threshold resummation beyond NLL that is valid across the entire phase space.
At variance with~\citere{Li:2016axz}, we found that the csoft function in regime 2 is not the same as the one in~\citere{Procura:2014cba}. This does not affect any of their other results, since they never use the expression obtained in~\citere{Procura:2014cba}.

\bibliography{pt_threshold}
 
\end{document}